\documentclass[prd,nofootinbib,twocolumn]{revtex4}
\usepackage{graphicx}
\usepackage{latexsym}

\def\be{\begin{equation}}
\def\ee{\end{equation}}
\def\bea{\begin{eqnarray}}
\def\eea{\end{eqnarray}}

\begin{document}

\title{Constraints on Dark Energy Parameters from Correlations of CMB with LSS}

\author{Hong Li${}^{a,b}$}
\author{Jun-Qing Xia${}^{c}$}

\affiliation{${}^a$Institute of High Energy Physics, Chinese Academy
of Science, P.O.Box 918-4, Beijing 100049, P.R.China}

\affiliation{${}^b$Theoretical Physics Center for Science Facilities
(TPCSF), Chinese Academy of Science, P.R.China}

\affiliation{${}^c$Scuola Internazionale Superiore di Studi
Avanzati, Via Beirut 2-4, I-34014 Trieste, Italy}

\begin{abstract}

In this paper, we combine the the latest observational data,
including the WMAP five-year data (WMAP5), the baryon acoustic
oscillations (BAO) and type Ia supernovae (SN) ``union" compilation,
and use the Markov Chain Monte Carlo method to determine the dark
energy parameters. We pay particular attention to the Integrated
Sache-Wolfe (ISW) data from the cross-correlations of cosmic
microwave background (CMB) and large scale structure (LSS). In the
$\Lambda$CDM model, we find that the ISW data, as a complement to
the WMAP data, could significantly improve the constraint of
curvature $\Omega_k$. We also check the improvement of constraints
from the new prior on the Hubble constant and find this new prior
could improve the constraint of $\Omega_k$ by a factor of 2.
Finally, we study the dynamical evolving EoS of dark energy from the
current observational data. Based on the dynamical dark energy
model, parameterizing as $w(a)=w_0+w_a(1-a)$, we find that the
$\Lambda$CDM model remains a good fit to the current data. When
taking into account the ISW data, the error bars of $w_0$ and $w_a$
could be shrunk slightly. Current constraints on the dynamical dark
energy model are not conclusive. The future precision measurements
are needed.

\end{abstract}

%\pacs{98.80.Es; 98.80.Cq}

\maketitle

%Introduction==========================================================

\section{Introduction}
\label{Int}
%Introduction of DE
Unveiling the origin of the current accelerating expansion of our
Universe is a big challenge for the modern cosmology either
theoretically or observationally. The origin source which drives the
expansion could be attributed to a mysterious budget, dark energy.
Thus, the nature of dark energy is one of the biggest unsolved
problems in modern physics and has been extensively investigated in
recent years.

%about isw
The measurements of CMB \cite{WMAP5GF1,WMAP5GF2,WMAP5Other}, LSS
surveys \cite{SDSS,2df} and SN \cite{Union,cfa} have provided a lot
of high quality data at present. These data have been widely used to
constrain various cosmological models. However, one should keep in
mind that the degeneracies of cosmological parameters generally
exist in almost all cosmological observations, i.e., they are not
sensitive to single parameters but to some specific combinations of
them. These degeneracies could weaken constraints on the
cosmological parameters. It is therefore highly necessary to combine
different probes to break parameter degeneracies so as to achieve
tight constraints. Furthermore, different observations are affected
by different systematic errors, and it is thus helpful to reduce
potential biases by combining different probes.

 One of the useful complementary probe is the late-time ISW effect
\cite{sachswolfe67}. This ISW effect is produced by the CMB photons
passing through the time-evolving gravitational potential well, when
dark energy or curvature becomes important at later times.
Therefore, the ISW effect provides a promising probe for studying
the acceleration mechanism of our universe, especially for the dark
energy and the curvature of Universe. Cross correlating CMB with
tracers of LSS surveys for detecting the ISW effect
\cite{Crittenden:1995ak} has been widely investigated in the
literature
\cite{Boughn:1997vs,Zaldarriaga:1998te,Hu:2001kj,Song:2002sg,
Hu:2001fb,Hu:2001tn,Afshordi:2003xu,Gaztanaga:2004sk,
Vielva:2004zg,Pietrobon:2006gh,McEwen:2006my,Giannantonio:2006du,
Rassat:2006kq,Ho:2008bz,Granett:2008ju,Xia:2009dr,Pogosian:2005ez}.

%outline of the paper
In this paper, we will present the constraints on various
cosmological models from the current observations, including the
WMAP5 data, SN ``union" compilation, and recently released BAO data
from SDSS DR7, as well as the ISW data. The structured of the paper
is as following: in Section II we describe the method and the data
sets; in Section III we present our numerical results and
discussion; finally we give a summary and outlook in Section VI.

%Method and Current Observations=======================================

\begin{figure*}[htbp]
\begin{center}
\includegraphics[scale=0.7]{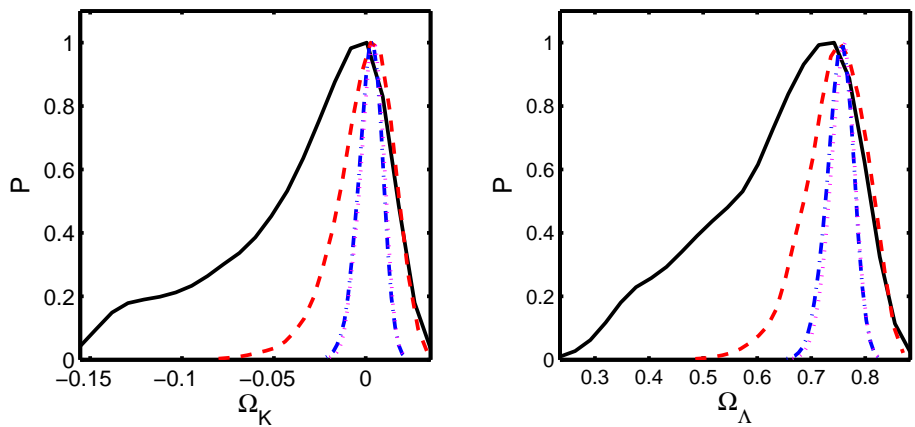}
\caption{One dimensional distributions of $\Omega_k$ and
$\Omega_{\Lambda}$ from different data combinations: WMAP5 (balck
solid lines), WMAP5+ISW (red dashed lines), WMAP5+HST (blue
dash-dotted lines), and WMAP5+HST+ISW (purple dotted lines).
\label{fig1}}
\end{center}
\end{figure*}

\section{Method and Data}
\label{Method}

%In our calculations, we perform a global analysis using the publicly
%available MCMC package CosmoMC\footnote{Available at:
%http://cosmologist.info/cosmomc/.} \cite{CosmoMC}. We have
%implemented the global fitting basing on two dark energy models:
For the parametrizations of dark energy models, we adopt
$\Lambda$CDM model and dynamical dark energy model with CPL
parametrization \cite{CPLparameter} as following:
\begin{equation}\label{cpl}
 w_{\rm DE}(a) = w_{0} + w_{a}(1-a)
\end{equation}
where $a=1/(1+z)$ is the scale factor and $w_{a}$ characterizes the
``running" of the equation-of-state (EoS) of dark energy. In the
$\Lambda$CDM model, $w_0=-1$ and $w_a=0$.

 For dark energy models whose EoS is not equal to $-1$ during the
evolution of Universe, the perturbation of dark energy will
inevitably exist. The perturbation of dark energy has no effect on
the geometric constraints. However, when including the CMB and LSS
data, the perturbation of dark energy should be fully considered,
because the late time ISW effect will be different significantly,
and will take an important signature on large angular scales of CMB
and the matter power spectrum of LSS \cite{Zhao:2005vj}. For
quintessence-like or phantom-like models, in which $w$ does not
cross the cosmological constant boundary, the perturbation of dark
energy is well defined. However, when $w$ crosses $-1$, one is
encountered with the divergence problem for perturbations of dark
energy at $w=-1$. In order to solve this problem in the global
analysis, we introduce a small positive constant $\epsilon$ to
divide the full range of the allowed value of the EoS $w$ into three
parts: 1) $ w
> -1 + \epsilon$; 2) $-1 - \epsilon \leq w  \leq-1 + \epsilon$; and
 3) $w < -1 -\epsilon $.

Working in the conformal Newtonian gauge, the perturbations of DE
can be described by \bea
    \dot\delta&=&-(1+w)(\theta-3\dot{\Phi})
    -3\mathcal{H}(c_{s}^2-w)\delta~~, \label{dotdelta}\\
\dot\theta&=&-\mathcal{H}(1-3w)\theta-\frac{\dot{w}}{1+w}\theta
    +k^{2}(\frac{c_{s}^2\delta}{{1+w}}+ \Psi)~~ . \label{dottheta}
\eea Neglecting the entropy perturbation, for the regions 1) and 3),
the EOS does not across $-1$ and the perturbation is well defined by
solving Eqs.(\ref{dotdelta},\ref{dottheta}). For the case 2), the
perturbation of energy density $\delta$ and divergence of velocity,
$\theta$, and the derivatives of $\delta$ and $\theta$ are finite
and continuous for the realistic dark energy models. However for the
perturbations of the parameterizations, there is clearly a
divergence. In our analysis for such a regime, we match the
perturbations in region 2) to the regions 1) and 3) at the boundary
and set $\dot{\delta}=0$ and $\dot{\theta}=0$. In our numerical
calculations we limit the range to be $|\Delta w = \epsilon
|<10^{-4}$ and find our method to be a very good approximation to
the multi-field dark energy model. More detailed treatments can be
found in Ref.\cite{Zhao:2005vj}.

The publicly available Markov Chain Monte Carlo (MCMC) package
CosmoMC\footnote{Available at:
http://cosmologist.info/cosmomc/.}\cite{CosmoMC} is employed in our
global fitting, and modifications have been made to include dark
energy perturbations, and to suit the dark energy models which we
study \cite{Zhao:2005vj,Xia:2005ge}. Furthermore, we assume purely
adiabatic initial conditions in our calculations.

Our most general theory parameter space vector is: %%
\begin{equation} \label{parameter}
{\bf P} \, \equiv \, (\omega_{b}, \omega_{c},
 \Theta_{s}, \tau, w_0, w_a, \Omega_k, n_{s}, A_{s}, c^2_s),
\end{equation}
 where $\omega_{b}\equiv\Omega_{b}h^{2}$ and
$\omega_{c}\equiv\Omega_{c}h^{2}$, in which $\Omega_{b}$ and
$\Omega_{c}$ are the physical baryon and cold dark matter densities
relative to the critical density, $\Omega_k$ is the spatial
curvature and satisfies $\Omega_k+\Omega_m+\Omega_{\Lambda}=1$,
$\Theta_{s}$ is the ratio (multiplied by 100) of the sound horizon
to the angular diameter distance at decoupling, $\tau$ is the
optical depth to re-ionization, $w_0$ and $w_a$ are the parameters
of dark energy EoS given by Eq.(\ref{cpl}), $A_{s}$ and $n_{s}$
characterize the power spectrum of primordial scalar perturbations,
$c^2_s$ is the sound speed of dark energy. For the pivot scale we
set $k_{s0}=0.05$Mpc$^{-1}$.

In the computation of CMB we have included the WMAP5 temperature and
polarization power spectra with the routine for computing the
likelihood supplied by the WMAP team\footnote{Available at the
LAMBDA website: http://lambda.gsfc.nasa.gov/.}. We also combine the
distance measurements from BAO and SNIa. For the BAO information, we
use the newly released gaussian priors on the distance ratios,
$r_s/D_v(z)=0.1905\pm0.0061$ at $z=0.2$ and
$r_s/D_v(z)=0.1097\pm0.0036$ at $z=0.35$, with a correlation
coefficient of $0.337$, which were measured from the power spectrum
for the distribution of the spectroscopic Sloan Digital Sky Survey
(SDSS) Data Release 7 (DR7) galaxy sample and the 2-degree Field
Galaxy Redshift Survey (2dFGRS) data\cite{Percival:2009xn}. For the
calculation of the likelihood from supernova, we use are the
``Union" compilation (307 sample) \cite{Union} and have marginalized
over the nuisance parameter \cite{SNMethod}.

For the ISW data, we have included the package for calculating the
ISW likelihood function provided by Ref.\cite{Ho:2008bz}, which
contain a $3.7\,\sigma$ detection of ISW by cross correlating WMAP
data with the LSS data sets of 2-Micron All Sky Survey sample, SDSS
photometric Luminous Red Galaxies, SDSS photometric quasars, and
NRAO VLA Sky Survey radio sources. In our calculations, we have
taken the total likelihood to be the products of the separate
likelihoods ${\bf \cal{L}}$ of CMB, BAO, SN and ISW. Furthermore, we
make use of the newly released HST prior on the Hubble constant,
which is the measurement of the Hubble parameter by the Near
Infrared Camera and Multi-Object Spectrometer (NICMOS) Camera 2 of
the Hubble Space Telescope (HST) and give $H_{0}\equiv
100$h~km~s$^{-1}$~Mpc$^{-1}$ by a Gaussian likelihood function
centered around $H_0=74.2$ and with a standard deviation
$\sigma=3.8$ \cite{Riess:2009pu}.

%Results===============================================================

\section{Numerical Results}
%Results(Dark Energy)==================================================

In this section we present our global fitting results of the
cosmological parameters determined from the latest observational
data.

%The numerical results from the global fitting of the constraints on
%cosmological parameters are listed in table I and table II. We have
%performed the global fitting in different dark energy models: in
%$\Lambda$CDM models with $w=-1$ we give the constraints on
%$\Omega_k$ and the sound speed $c_s$, and we also specially consider
%the constraints on dynamical dark energy model with the
%parametrization in EQ. (\ref{cpl}). We in particular compare the
%cases of the data combination with or without the ISW in order to
%see its effects.

\begin{table*} \hspace{-5mm}
TABLE I. Mean $1\,\sigma$ constraints on $\Omega_k$ and
$\Omega_{\Lambda}$ from different data combinations in the
$\Lambda$CDM model.
\begin{center} %\hspace{-15mm}
\begin{tabular}{c|c|c|c|c}
\hline\hline
&WMAP only & WMAP+ISW&WMAP+HST&WMAP+ISW+HST\\
\hline
$100\times\Omega_k$ &$-3.57\pm4.28$&$-0.182\pm1.61$&$0.226\pm0.65$&$0.304\pm0.630$  \\
\hline
$\Omega_{\Lambda}$ &$0.635\pm0.130$&$0.742\pm0.0566$&$0.752\pm0.0243$&$0.756\pm0.0228$  \\
\hline\hline
\end{tabular}
\end{center}
\end{table*}

In Table I we list the constraints on the dark energy density
$\Omega_\Lambda$ and the curvature $\Omega_k$ in the $\Lambda$CDM
model from different data combinations. Due to the well-known
degeneracy between $\Omega_\Lambda$ and $\Omega_k$, we obtain a weak
constraint on the curvature $\Omega_k$ from WMAP5 data only,
$\Omega_k=-0.036\pm0.043$ ($68\%$C.L.), as shown in Fig.\ref{fig1}.
Our universe is very close to flatness, which is consistent with the
prediction of inflation paradigm. However, this degeneracy could be
broken by adding other different cosmological data, such as the
large scale structure and supernovae data. When we add the ISW data
into the analysis, we can find that the combined constraint from
WMAP5+ISW is significantly improved over using WMAP5 alone, namely
the $68\%$ interval is $\Omega_k=-0.002\pm0.016$. The constraints of
$\Omega_k$ and $\Omega_{\Lambda}$ improve by a factor of $2.6$ and
$2.2$, respectively. The ISW data give the remarkable complementary
effect on WMAP5 data in constraining cosmological parameters and
break the degeneracy between $\Omega_k$ and $\Omega_{\Lambda}$.

In Fig.\ref{fig1} we also plot the one dimensional posterior
distribution of $\Omega_k$ and $\Omega_{\Lambda}$ from WMAP5+HST
prior. Here, we consider the new released HST prior,
$h=0.742\pm0.038$ ($1\,\sigma$). We find that this new HST prior
could play an important role in constraining cosmological parameter
and give much tighter constraints on cosmological parameters,
$\Omega_k=0.002\pm0.007$ ($1\,\sigma$). The error bar has been
shrunk by a factor of $6.6$ when compared to the constraint from
WMAP5 alone. The $95\%$ intervals of $\Omega_k$ and $\Omega_\Lambda$
are $-0.012<\Omega_k<0.014$ and $0.70<\Omega_\Lambda<0.80$. We also
do the calculations using the old HST prior, $h=0.72\pm0.08$
\cite{oldHST}. Using WMAP5+HST old prior, we obtain that the
$2\,\sigma$ constraints on $\Omega_k$ and $\Omega_{\Lambda}$ are
$-0.045<\Omega_k<0.015$ and $0.59<\Omega_\Lambda<0.80$,
respectively. The constraints significantly improve by a factor of
$2$. Furthermore, when combining WMAP5, ISW and HST together, the
constraint on $\Omega_k$ will improve further:
$\Omega_k=0.003\pm0.006$ at $68\%$ confidence level.

\begin{figure}[t]
\begin{center}
\includegraphics[scale=0.45]{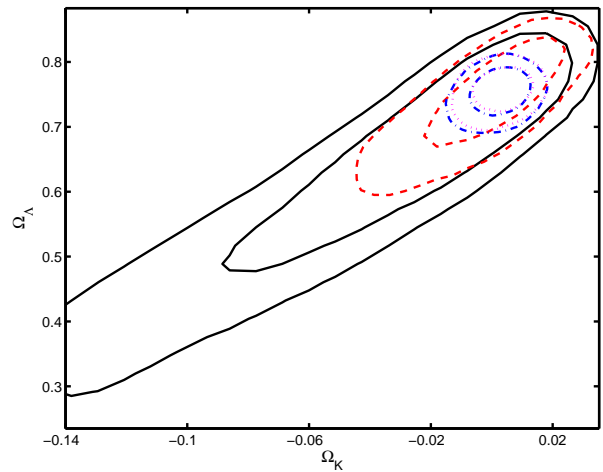}
\caption{Two dimensional contours of $\Omega_k$ and
$\Omega_{\Lambda}$ from different data combinations within the
$\Lambda$CDM model: WMAP5 (balck solid lines), WMAP5+ISW (red dashed
lines), WMAP5+HST (blue dash-dotted lines), and WMAP5+HST+ISW
(purple dotted lines). \label{fig2}}
\end{center}
\end{figure}

In Fig.\ref{fig2} we present the two dimensional contours of
$\Omega_k$ and $\Omega_{\Lambda}$ from different data combinations
in the $\Lambda$CDM model. We find that $\Omega_k$ and
$\Omega_{\Lambda}$ are highly correlated with each other. The ISW
data and the new HST prior are very helpful in breaking such
degeneracy.

\begin{figure}[t]
\begin{center}
\includegraphics[scale=0.45]{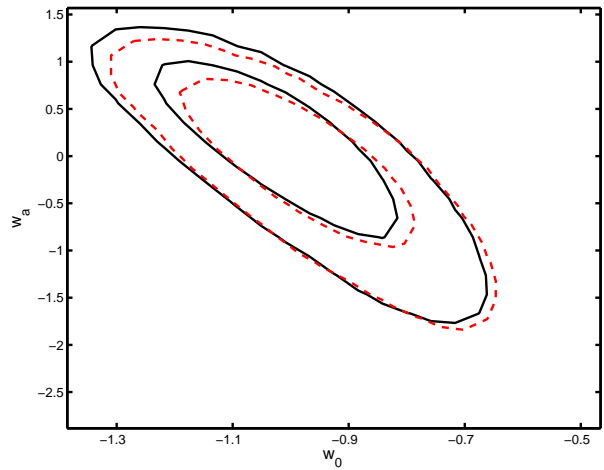}
\caption{Two dimension contours of the EoS parameters of dark energy
$w_0$ and $w_a$. The black solid lines are given by WMAP5+SN+BAO and
the red dashed lines are given by WMAP5+SN+BAO+ISW,
respectively.\label{fig3}}
\end{center}
\end{figure}

\begin{table} \hspace{-5mm}
TABLE II. Mean $1\,\sigma$ constraints on $w_0$, $w_a$ and
$\Omega_{\Lambda}$ from different data combinations in the dynamical
dark energy model.
\begin{center} %\hspace{-15mm}
\begin{tabular}{c|c|c}
\hline\hline
&WMAP+SN+BAO& WMAP+SN+BAO+ISW \\
\hline
$w_0$ &$-1.002\pm0.136$&$-0.970\pm0.130$  \\
\hline
$w_a$ &$0.0096\pm0.64$&$-0.134\pm0.612$  \\
\hline
$\Omega_{\Lambda}$ &$0.732\pm0.0149$&$0.734\pm0.0144$  \\
\hline\hline
\end{tabular}
\end{center}
\end{table}

We now present the constraints on dark energy parameters. In
Ref.\cite{Ho:2008bz}, they considered the constraints on dark energy
model with constant EoS, and found that the ISW data could modestly
improve constraint of dark energy EoS. As we know, the variation of
dark energy EoS could also affect the evolution of gravitational
potential, and the ISW effect consequently. It would be worth
checking the capabilities of ISW data on the constraint of
time-varying dark energy EoS. In Fig.\ref{fig3} we plot the two
dimensional contours of $w_0$ and $w_a$ from different data
combinations. The black solid lines are given by fitting with the
combination of WMAP5 temperature and polarization power spectra,
supernovae data, as well as the newly released BAO data. Due to the
limits of the precisions of observational data, the variance of
$w_0$ and $w_a$ are still large, namely the $95\%$ constraints on
$w_0$ and $w_a$ are $-1.25<w_0<-0.72$ and $-1.47<w_a<1.00$
respectively, which are listed in Table II. This result implies that
the dynamical dark energy models are not excluded and the current
data cannot distinguish different dark energy models decisively. The
$\Lambda$CDM model, however, is still a good fit right now.

We also include the ISW data into the calculations. From the red
dashed lines in Fig.\ref{fig3}, we can find that the constraints on
the dark energy EoS parameters slightly improve from
WMAP5+SN+BAO+ISW combination. The $95\%$ intervals are
$-1.21<w_0<-0.71$ and $-1.52<w_a<0.86$. The ISW data help tighten
the lower limit on $w_0$ and the upper limit on $w_a$, but not
significantly. These is due to the constraining power of SNIa and
BAO, while the current ISW data are still not constraining enough.
However, the ISW data could be a useful probe for constraining other
dark energy models, such as the early dark energy model
\cite{Xia:2009dr}. And they also found that the future ISW data can
improve the constraints on the most important cosmological
parameters by a factor $\sim1.5$.

Again, we also compare the constraints on dark energy parameters
between from the new HST prior and from the old one. When we combine
WMAP5, SN, BAO, as well as the old HST prior, we obtain that the
$2\,\sigma$ constraints on $w_0$ and $w_a$ are $-1.23<w_0<-0.7$ and
$-1.49<w_a<0.94$, which is slightly weaker than those from the new
HST prior.

If dark energy is not the cosmological constant, we should consider
the dark energy perturbations in our calculations. In the framework
of the linear perturbations theory, besides the EoS of dark energy,
the dark energy perturbations can also be characterized by the sound
speed, $c_s^2\equiv \delta p_\Lambda/\delta\rho_\Lambda$. The sound
speed of dark energy affects the evolution of perturbations, and
leaves the signatures on the CMB power spectrum
\cite{Weller:2003hw,Hu:2004yd,Xia:2007km}. Thus, in the literature
constraining on the sound speed $c^2_s$ from different observational
data has been widely investigated (e.g.
Refs.\cite{DeDeo:2003te,Erickson:2001bq,Xia:2007km}).

\begin{figure}[t]
\begin{center}
\includegraphics[scale=0.45]{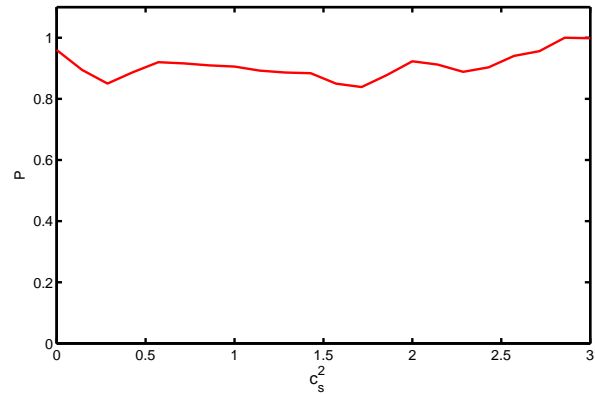}
\caption{One dimensional constraints on the sound speed $c_s^2$ from
WMAP5+SN+BAO+ISW combination in the dynamical dark energy model.
\label{figcs}}
\end{center}
\end{figure}

Here, we also constrain the dark energy sound speed from current
data in the dynamical dark energy model. In Fig.\ref{figcs}, we show
the one dimensional constraint on the dark energy sound speed
$c_s^2$ from the combination of WMAP5 temperature and polarization
power spectra, SNIa and BAO, as well as the ISW data. We can find
that the constraint on the dark energy sound speed $c^2_s$ are still
very weak. The current observational data are still not accurate
enough.

%Summary===============================================================

\section{Summary}
\label{Sum}

In this paper, we study the constraints on cosmological parameters
from the recently released CMB, BAO and SNIa data. Here, we pay
particular attention to the current ISW data which is the cross
correlations of CMB with LSS surveys. In the $\Lambda$CDM model, the
ISW data and the new HST prior are very helpful to break the
degeneracy between $\Omega_k$ and $\Omega_\Lambda$. The constraints
on $\Omega_k$ and $\Omega_\Lambda$ significantly improve, when
compared to the constraints from WMAP5 alone.

More importantly, we consider the constraints on the dark energy
parameters in the CPL dark energy model. Here, we fully include the
perturbations of dark energy. This result implies that the dynamical
dark energy models are not excluded, the $\Lambda$CDM model,
however, is still a good fit right now.  We find that the ISW data
could give slight improvement of the constraints on CPL dark energy
model. But this does not mean that the ISW data can not constrain
the dynamical dark energy models efficiently. Actually, it should be
useful for constraining the dynamical dark energy models whose EoS
$w(z)$ deviate from the cosmological constant boundary obviously,
for example, the early dark energy models \cite{Xia:2009dr}, or the
dark energy model with its EoS $w(z)$ transits sharply during its
evolution. Besides the dark energy models, the ISW data could be
also helpful for testing the modified gravity theories
\cite{Jain:2007yk,Afshordi:2004kz,Zhang:2005vt,Schmidt:2007vj},
massive neutrinos \cite{Lesgourgues:2007ix}, the primordial
non-gaussianity \cite{slosar,xia10}, and so on.

Furthermore, we compare the constraints between from the new HST
prior and from the old one. One can see that the new HST prior gives
the tighter constraints on the cosmological parameters. Finally, we
check the capability of current observational data to constrain the
dark energy sound speed $c_s^2$. We find that the sound speed is
weakly constrained by current observations, and thus futuristic
precision measurements of the CMB on a very large angular scale (low
multipoles) are necessary.

%Acknowledgments=======================================================

\section*{Acknowledgements}

We acknowledge the use of the Legacy Archive for Microwave
Background Data Analysis (LAMBDA). Support for LAMBDA is provided by
the NASA Office of Space Science. Our numerical analysis was
performed on the MagicCube of Shanghai Supercomputer Center (SSC).
We thank Yi-Fu Cai for helpful discussions. This work is supported
in part by the National Natural Science Foundation of China under
Grant No. 10803001, 973 Program under Grant No.2010CB833000 and the
Youth Foundation of the institute of high energy physics under Grant
Nos. H95461N.

%End===================================================================

\end{document}